\documentclass[12pt]{article}
\usepackage{epsfig}
\usepackage{cite}
\usepackage{mcite}
\usepackage{url}
\usepackage{array,tabularx,epsfig,mathrsfs,graphicx,rotating}
\usepackage{ifthen}
\usepackage{fancybox}
\usepackage{amsfonts}

\usepackage[pagewise]{lineno}

\newcommand{\beq}{\begin{equation}}
\newcommand{\eeq}{\end{equation}}

\chardef\til=126

\newcommand{\LMJ}{\vec{L}_{MJ}}
\newcommand{\LMI}{\vec{L}_{MI}}
\newcommand{\LMJA}{\vec{L}_{MJ}^{(1)}}
\newcommand{\LMJB}{\vec{L}_{MJ}^{(2)}}
\newcommand{\LMIA}{\vec{L}_{MI}^{(1)}}
\newcommand{\LMIB}{\vec{L}_{MI}^{(2)}}

\begin{document}

\clearpage
\pagestyle{empty}
\setcounter{footnote}{0}\setcounter{page}{0}%
\thispagestyle{empty}\pagestyle{plain}\pagenumbering{arabic}%

\hfill  ANL-HEP-PR-10-47

\hfill September 10, 2010

\vspace{2.0cm}

\begin{center}

{\Large\bf
New approach for jet-shape identification of TeV-scale  particles 
at the LHC
\\[-1cm] }

\vspace{2.5cm}

{\large S.V.~Chekanov, C.~Levy\footnote[1]{also affiliated with the Physics Department, Northeastern University
      110 Forsyth St., 111 Dana Research Center
      Boston, MA. 02115}, J.~Proudfoot, R.Yoshida  

\vspace{0.5cm}
\itemsep=-1mm
\normalsize
\small
HEP Division, Argonne National Laboratory,
9700 S.Cass Avenue, \\ 
Argonne, IL 60439
USA
}

\normalsize
\vspace{1.0cm}


\vspace{0.5cm}
\begin{abstract}
A new approach to jet-shape identification based on linear regression
is discussed.
It is designed for searches   
for new particles at the TeV scale  decaying hadronically
with strongly collimated jets.
We illustrate the method using a Monte Carlo
simulation for $pp$ collisions at the LHC
with the goal to reduce the contribution
of QCD-induced events. We focus on 
a rather generic example $X\to t\bar{t}\to hadrons$,
with $X$ being a heavy particle, 
but the approach is well suited for reconstruction of other decay channels
characterized by a cascade decay of known states.  
\end{abstract}

\end{center}

\newpage
\setcounter{page}{1}


\section{Introduction}

A promising path for discoveries of TeV-scale particles to be produced 
at the LHC is through model-independent searches in which events can be
classified in exclusive classes according to 
the number of identified high-$p_T$ objects (jets and leptons).
In the case of heavy particles, with masses close to the TeV scale, 
the decay products undergo a significant Lorentz boost, and this 
leads to their partial or complete overlap. 
In the case of jets, this closes the opportunity
of bump hunting in invariant-mass
spectra using individual jets since the event signatures will be indistinguishable from 
those of the standard QCD-induced events.
 
Jet shapes are often discussed as a useful tool to 
disentangle events induced by the standard QCD processes from those containing
jets as the results of decay products of TeV-scale particles.  
They are expected to be useful
in reduction of the overwhelming rate
of conventional QCD jets, thus opening the path to a direct observation of new  states.
\cite{Agashe:2006hk,Lillie:2007yh,Butterworth:2007ke,Almeida:2008tp,Almeida:2008yp,
Kaplan:2008ie,Brooijmans:2008,Butterworth:2009qa,Ellis:2009su,ATL-PHYS-PUB-2009-081,CMS-PAS-JME-09-001,Chekanov:2010vc,Almeida:2010pa,Hackstein:2010wk}.  

In this paper we extend the studies of jet shapes presented in \cite{Chekanov:2010vc}
for the generic decay channel $X\to t\bar{t}\to hadrons$,
with $X$ being a heavy particle with  
a mass close to the TeV scale. 
It is assumed that the mass of a particle $X$ is so heavy that the top quarks
will form two energy deposits in cones around the top-quark directions. 
Thus, given the  finite spacial resolution of a detector, the decay products of top quarks
will be seen as monojets. It is hoped that shapes of such monojets will be different
from those of the standard QCD jets, with direct implications for experimental searches of heavy particles.

While the approach presented in \cite{Chekanov:2010vc}
was mainly based on two jet-shape variables, jet width and eccentricity derived
using the principle-component analysis of jet constituents, in this paper
we will propose a more intuitive approach which provides a significantly
larger number of jet-shape characteristics.
In fact, the approach proposed in this paper goes beyond the jet-shape 
identification and deals with a general problem
of a dimensionality reduction, i.e. 
how to reduce the amount of information in the original multi-dimensional data keeping only
a few parameters which catch the most basic spacial features of the original data.
In the case of the jet-shape studies, we are interested  not only in the extent of elongation   
of the jet shape characterized by the eccentricity, but also in a 
degrees of skewness of jet shapes  which cannot be easily estimated using the techniques discussed before \cite{Chekanov:2010vc}.

\section{Jet shapes and jet masses} 
\label{sec1}

The jet-shape analysis performed in Ref.~\cite{Chekanov:2010vc} included mass
cuts and cuts on the jet shapes (the jet width and the eccentricity).
The cut on the jet masses has by far the most rejection
for QCD-jets. Indeed, the channel  $X\to t\bar{t}\to hadrons$ features
two monojets, each of which has a mass close to the top mass. Thus,
selecting events with two jets with masses above some cut close to the top mass,
one can reject a significant fraction of the standard QCD jets which have an exponentially
decreasing mass spectra, unlike monojets from the top decays.

It has already been shown ~\cite{Chekanov:2010vc} that 
there is a strong positive correlation between the jet mass
and the jet width, thus applying cuts on jet mass and jet width at the same time
may lead to unoptimized rejection factors. Therefore, in this paper we   
take a different approach and apply the mass cut before considering jet-shape variables.
Table~\ref{table1} shows the rejection factors after using the jet-mass cuts on two leading jets
with  $p_T>500$ GeV. 
The analysis was performed using the PYTHIA Monte Carlo model \cite{Sjostrand:2006za}
included in the RunMC package~\cite{runmc}
which interfaces FORTRAN Monte Carlo models with ROOT~\cite{root} and other C++ libraries.
Jets and their shapes were reconstructed using the FastJet package~\cite{fastjet}.
The jets were reconstructed with the anti-$k_T$ algorithm \cite{Cacciari:2008gp} 
with  a distance parameter of 0.6. 
Currently, this jet algorithm is 
the default for jet reconstruction at the ATLAS and CMS experiments.
We simulated heavy-particle decays using $Z'$ bosons as they are included in the
PYTHIA model, forcing such states to decay to $t\bar{t}$ pairs.
Both top quarks were set to decay  hadronically.
The PYTHIA parameters were set to the default ATLAS parameters tuned
to describe multiple interactions \cite{Moraes:2009zz}.
The events were first generated and stored for easy processing.

Table~\ref{table1} shows that the 
rejection factor after the jet-mass cut varies from $\sim 10$ 
to $\sim 400$ 
for the standard QCD events, while the mass cuts have a small effect on the events with TeV-scale
particles, leading to a rejection between 1 and 2.3.  For example, 
for the 70 GeV mass cut, the rejection factor for QCD events
is roughly 9.9, while it is only a factor of $1.1$ for the 2 TeV signal events.
Therefore, the ratio of the rejection factors for inclusive QCD and events 
with heavy states
is about 9. 

For the analysis of jet shapes in the next section, we will consider monojets with approximately
similar masses, close to the top mass. We have chosen  the jet-mass range  $140<M(jet)<300$ GeV.
With such a tight mass constraint, the jet shapes should mainly reflect the spacial
distribution of jet constituents for kinematically similar jets (with similar transverse momenta and jet masses).
Keeping this mass range in mind, we will attempt to find
differences in jet shapes for QCD events and events originating from
$X\to t\bar{t}\to hadrons$.

\begin{table}
\begin{center}
\begin{tabular}{|c|c|c|c|}
\hline
      & $M(jet) >70$  GeV & $M(jet)>100$ GeV & $M(jet) >140$ GeV  \\
\hline
$pp$   & 9.9    & 48 &  380  \\
2 TeV  & 1.1    & 1.4  & 2.3  \\
3 TeV  & 1.05   & 1.2  & 1.6  \\
4 TeV  & 1.06   & 1.1  & 1.3  \\
\hline
\end{tabular}
\caption{
Event rejection factors for inclusive $pp$ events and events with TeV-scale particles
$Z' \to t\bar{t} \to 6 q$, where $Z'$ has a mass ranging from 2 TeV to  4 TeV.
The rejection factors were calculated by applying the cuts 
$70$, $100$ or $140$ GeV on the mass of two leading in $p_T$ jets. } 
\label{table1}
\end{center}
\end{table}

\section{Jet shapes using linear-regression analysis}
\label{sec2}

To characterize jet shapes in two-dimensions, say in $\eta$ (pseudorapidity) and $\phi$ (azimuthal angle),
we propose a new approach which is significantly different from 
the principle-component analysis considered in Ref.~\cite{Chekanov:2010vc}.
Being more intuitive, this method will also allow us to construct a significant number of jet-shape
variables sensitive to the jet size in the transverse and longitudinal directions,
as well as  the degree to which the jet constituents form skewed shapes.

Let us consider a jet constituent (hadron, calorimeter cluster, etc.) 
defined by its position in $\eta$ (pseudorapidity) and $\phi$ (azimuthal angle)
with respect to the beam-line and interaction-point, as well as by its energy $e$. 
In this case, each particle is
represented by a point ($\eta$ and $\phi$) and a weight ($e$), 
making the shape effectively three-dimensional. If
it is assumed that the jet in this phase space is a conic section (roughly elliptical), then we can define
several shape variables, including the major axis length, minor axis length, ellipse eccentricity, and others
(to be discussed below). The first task is thus to define the axes and lengths of the ellipse.

\begin{figure}[htp]
\begin{center}

\begin{minipage}[b]{0.45\linewidth}
\centering
\includegraphics[scale=0.3]{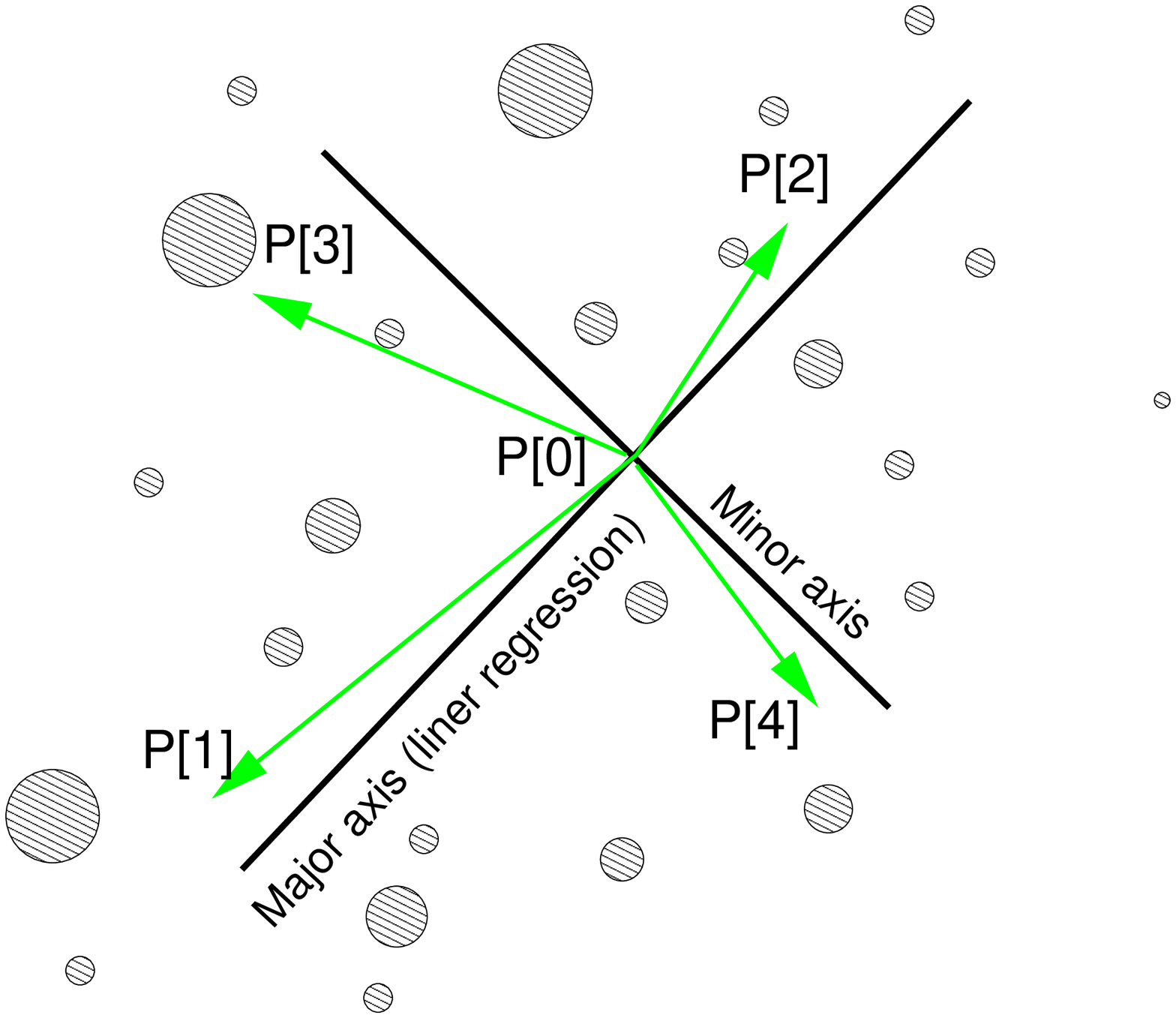}
\end{minipage}
\begin{minipage}[b]{0.45\linewidth}
\centering
\includegraphics[scale=0.3]{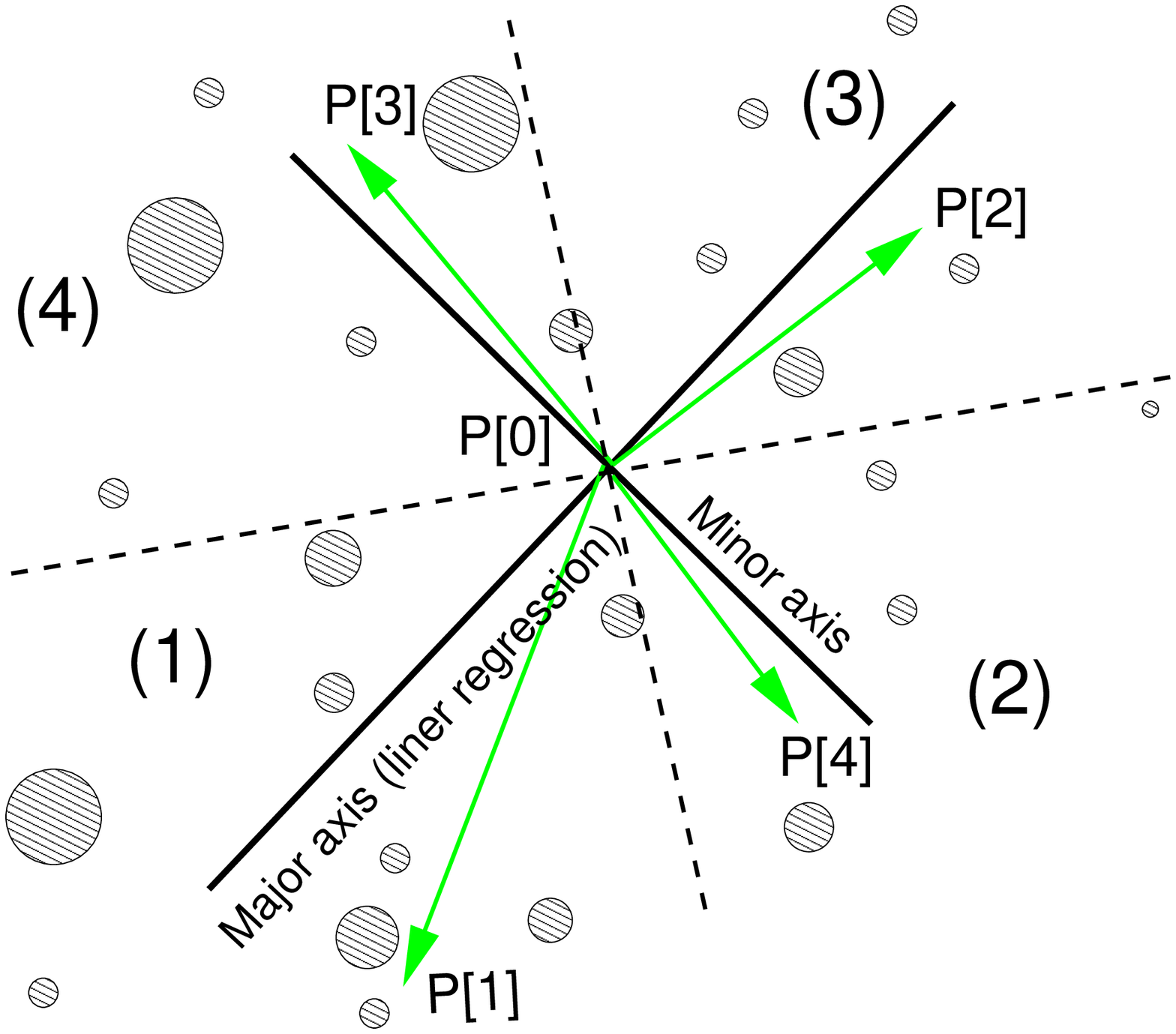}
\end{minipage}
\end{center}
\caption{Sketch of two approaches for analyzing an approximately elliptical composite object.
Each point represents  a jet constituent, with the size 
representing its weight. The geometric 
major axis is defined by an unweighted linear regression, the geometric minor is
by definition perpendicular to the major axis, through the geometric mean.
In the non-quadrant method (left figure), weighted centers P[N] ($N=1,\ldots ,4$) are defined for the regions
above or below the major and minor axes. In the quadrant-method (right figure), each weighted center P[N] 
is uniquely associated with one of four quadrants shown with dashed lines and denoted as (1)-(4). In both methods,
the weighted centers do not need to be located on the axes.  
}
\label{Sketch}
\end{figure}

First, a linear regression analysis in two-dimensions 
is performed to define the direction of geometrical elongation in $\eta$ and $\phi$.   
The linear regression was performed using the least squares approach by minimizing 
the sum of the squares of the vertical distances of the points from the line. 
At this stage, all data points 
are assumed to have exactly the same weights.
The linear regression defines the best-fit values of the slope and intercept of  
the major axis. 
In the calculations, a geometric mean of all constituents is found first  
(P[0] in Fig.~\ref{Sketch}) and then a linear regression is performed. 
The major axis is given by the fit, while
the minor axis is defined to be perpendicular to the major axis and passing
through the geometric mean.

With the axis-lines of the ellipse defined, the next step is to calculate the axis length. We identified
two main classes of length definition: the quadrant method and the non-quadrant method. We will
discuss each of these methods respectively.

\subsection{Quadrant Method (QM)} 
In the quadrant method (QM), the $\eta-\phi$ space is first divided into four quadrants centered at the
ellipse geometric center, each of which corresponds to one of the ellipse semi-axes (Fig.~\ref{Sketch} (right)).
This is done by
taking the major and minor axis lines (from linear regression) and rotating them by $45^o$, putting each
semi-axis in one of the quadrants,  as this shown using 
dashed lines in Fig.~\ref{Sketch}(right). The length of each semi-axis is defined by finding the
weighted center of each quadrant; that is, all constituent points are separated by the quadrant in which
they lie and the weighted mean of each quadrant is found independently, without consideration of  points
in other quadrants. 
Each data point is uniquely associated with each quadrant.
The length of the semi-axis is thus the length between the global geometric center
and the quadrant center. 

The semi-axes are sensitive to spacial positions of jet constituents in 
3D (where the third component, the constituent weight, is given by energy),  
since the geometrical axes are defined using unweighted regression, while
the  semi-axes are calculated using weights.

\subsection{Non-Quadrant Method (NQM)}
In the non-quadrant method (NQM), the major and minor axis themselves define the areas where the weighted means
are calculated;
the major axis-line defines two semi-planes (the part above and the part below), as
does the minor axis-line, see Fig.~\ref{Sketch}(left).
In this way, each point is in two of four semi-planes rather than a
single exclusive quadrant. The weighted means above and below the major axis-line are the weighted
centers defining the lengths of the minor semi-axes, while the means above and below the minor axis
define the lengths of the major semi-axes.

As example, the point P[3] in Fig.~\ref{Sketch}(left) shows a weighted mean of the area above the major axis,
while P[4] defines the same but for the plane below the major axis.
Similarly, [P1] and P[2] show the weighted means for the plane below and above the minor axis.
It is important to note that all four centers are defined using weights (i.e. jet constituent energies),
which increase the sensitivity to data in 3D.
The distances between the points P[1] and P[0] (P[2] and P[0]) define
major semi-axes. Analogously, the distances between the points [P3] and P[0] ([P4] and P[0])
define the minor semi-axes.

The main difference between the QM and NQM is different sensitivity to spacial topology:
Each point in the NQM contributes to both major and minor semi-axes.
Thus, the  NQM is more sensitive to a shape  of elongated distribution, since all
points from both sides of the major axis contribute to the (semi)-minor length.
For example, a shape with roughly the same width along the major axis
(a pencil-like shape) indicates a very small minor length. 

For the QM method, points are uniquely associated with each
quadrant and can contribute either to the minor or the major semi-axis. 
In the example with a pencil-like shape discussed above, only
 a small fraction of phase 
space close to the geometrical center can contribute to the minor axis.
Adding an extra point in the region of $45^o$ from one side of the major axis 
will have a strong impact on the minor semi-axis, without contribution
to the major semi-axis (unlike the NQM definition).

\subsection{Definition of Variables} 

The geometrical major and minor axes from the linear-regression analysis were only
necessary to define the regions with four positions  
of semi-axes which are used
for calculations of actual major and minor vectors and jet-shape variables based on these vectors.
Each variable
can be defined either in QM or NQM. 

\begin{itemize}

\item
Major length, $|\LMJ|$, a distance between major semi-axis centers (P[1] and P[2]) which defines
the size of longitudinal elongation (which includes 2D geometry and weights given by the 
energies of constituents).  
It can be composed into two semi-axes from each side of the minor axis.
By definition, $\LMJ=\LMJA-\LMJB$ and $|\LMJA|>|\LMJB|$,
where $|\LMJA|$ is the longest and $|\LMJB|$ shortest length of the semi-axis. 

\item
Minor length, $|\LMI|$, a distance between minor semi-axis centers (P[3] and P[4]) which
defines
the size of the transverse  elongation.
It can be decomposed into  two semi-axes from each side of the major axis.
By definition, $\LMI=\LMIA-\LMIB$ and $|\LMIA|>|\LMIB|$, 
where $\LMIA$ is the longest and $|\LMIB|$ the shortest length of the minor semi-axis.

\item
Eccentricity, $ECC$, defined as
$$
ECC=1-\frac{|\LMI|}{|\LMJ|} 
$$
with the range $[0,1]$.
This variable measures the degree to which the ellipse fails to be circular. $ECC=0$ is for a perfect circle,
and 1 for  an infinitely elongated object.
For the QM,  this parameter emphasizes the relative width of an elongated object due to
contributions of points closer to the geometrical center, while the same parameter in the
NQM is more sensitive to the contribution of points away from the center.    

\item
Major eccentricity, $ECC_{MJ}$:
$$
ECC_{MJ}=1-\frac{|\LMJB|}{|\LMJA|}, 
$$
where $|\LMJA|$ and $|\LMJB|$ are the lengths of the major  semi-axes as defined above. 
This is a measure of the degree to which the ellipse is skewed to one side of the minor axis. A
large value signifies a large difference between lengths of the major semi-axes.
For the QM, it is sensitive to skewness due to points close to the geometrical center,
while for the NQM it is more sensitive to points at the shape edges.

\item
Minor eccentricity, $ECC_{MI}$:
$$
ECC_{MI}=1-\frac{|\LMIB|}{|\LMIA|}, 
$$
where $|\LMIA|$ and $|\LMIB|$ are the lengths of the minor semi-axes. 
This is a measure of the degree to which the ellipse is 'skewed' to one side of the major direction. A
large value signifies a large difference between lengths of the minor semi-axes.
As before, the value of $ECC_{MI}$ is in the range [0,1].
The differences between the QM and NQM methods  are as for the $ECC_{MJ}$. 
\end{itemize}

The above variables can be defined using either the QM or NQM.  
As mentioned above, the QM  is sensitive to the asymmetries in the shape close to the geometrical center
of the entire distribution, 
while the NQM is more sensitive to asymmetry at the edges. 


\begin{figure}[ht]

\begin{minipage}[b]{0.45\linewidth}
\centering
\includegraphics[scale=0.3]{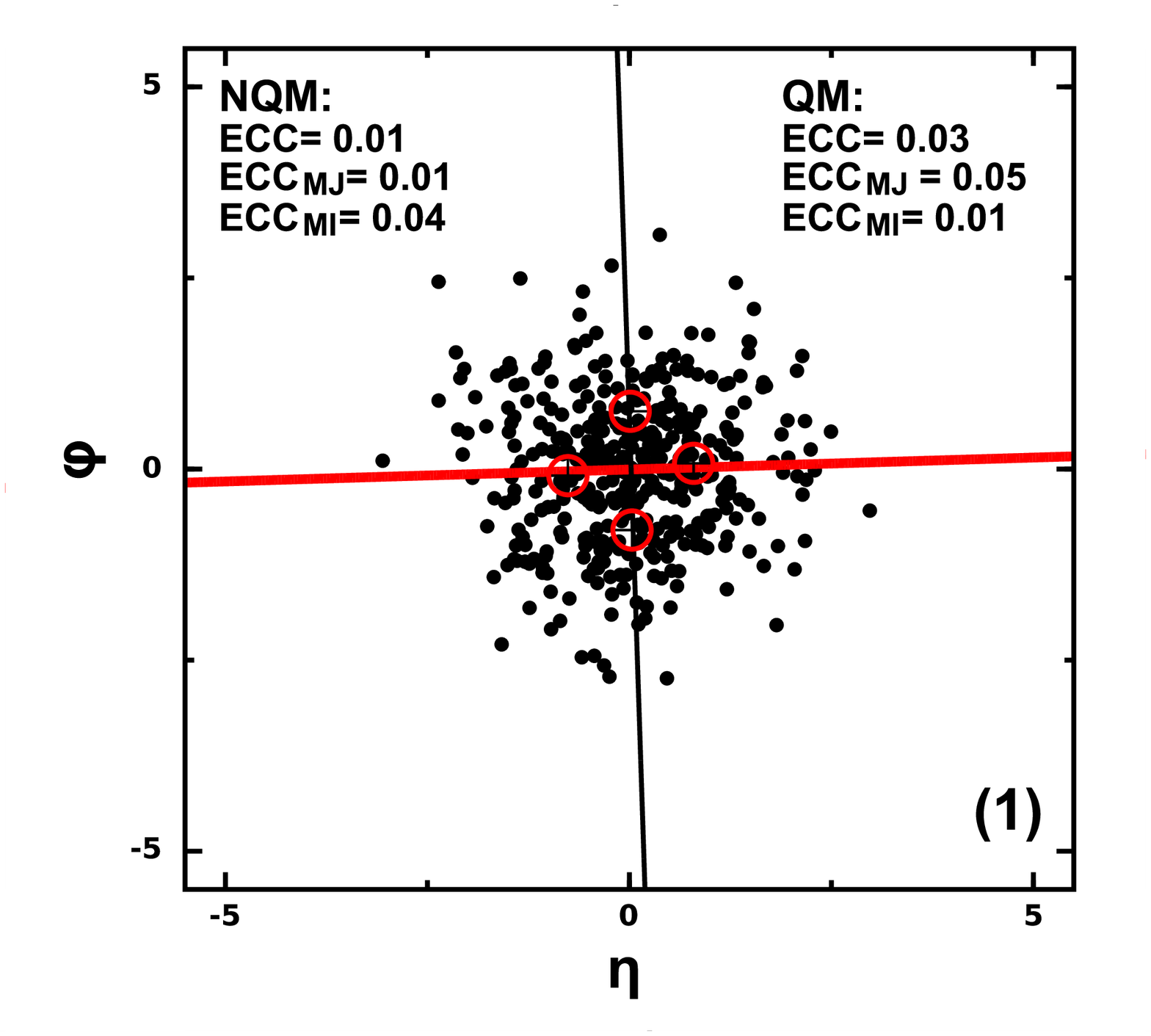}
\end{minipage}
\begin{minipage}[b]{0.45\linewidth}
\centering
\includegraphics[scale=0.3]{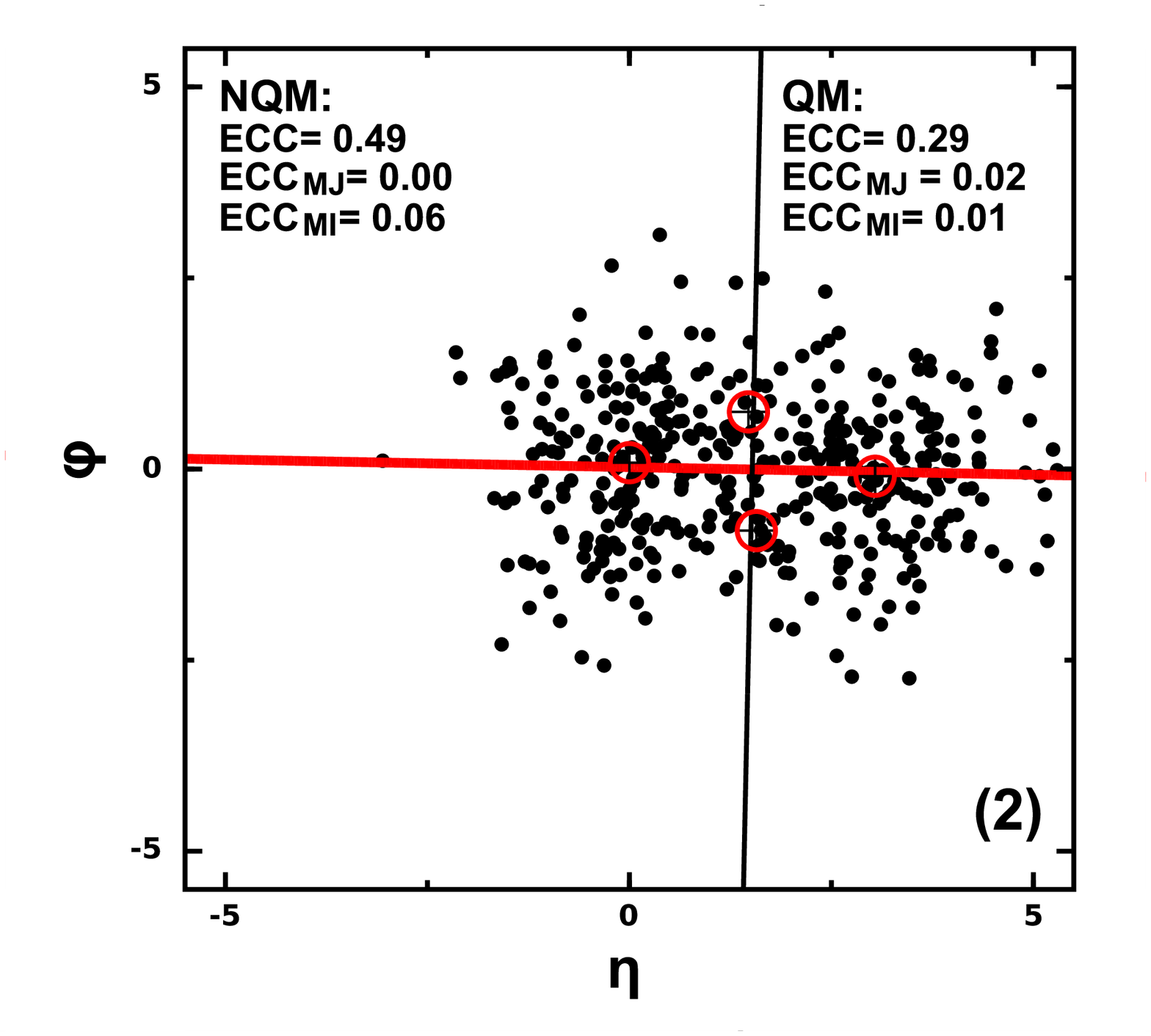}
\end{minipage}

\begin{minipage}[b]{0.45\linewidth}
\centering
\includegraphics[scale=0.3]{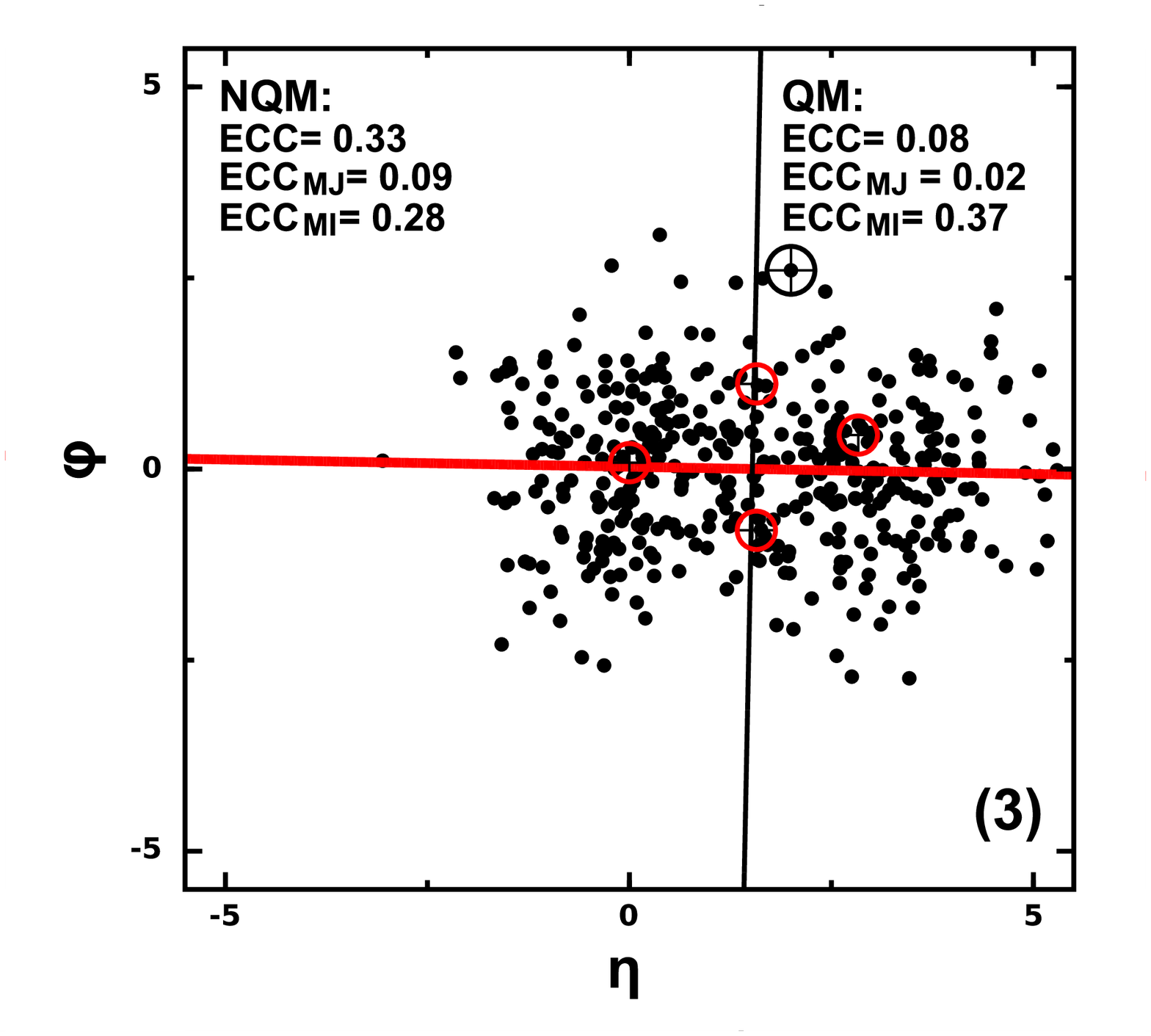}
\end{minipage}
\begin{minipage}[b]{0.45\linewidth}
\centering
\includegraphics[scale=0.3]{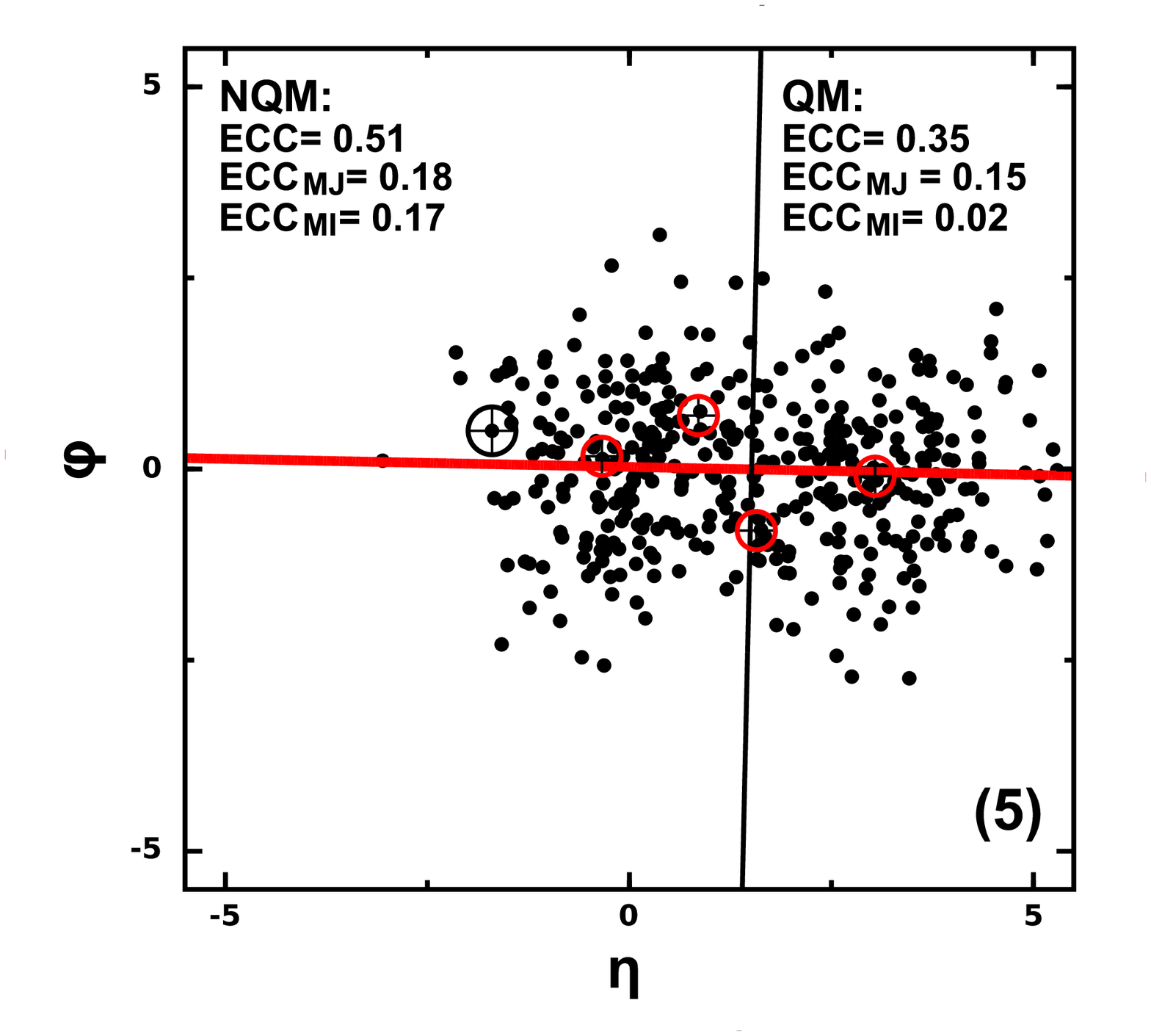}
\end{minipage}
\caption{Several examples illustrating the concept of the shape-variables in the linear-regression approach. 
400 points in $\eta$ and $\phi$ phase space are distributed randomly 
using two overlapping Gaussian distributions: (1) the mean positions
of the Gaussian distributions  are the same;  
(2) two Gaussian distributions are shifted by 3 (arbitrary) 
in $\eta$-direction; 
(3) the same as before, but a new point  
was added with the weight 50 (open circle); (4) the 
heavy-weight point was moved closer to the major axis.  
The thick (red) line shows the linear regression which defines the  major axis, 
while thin black line is the minor axis. The eccentricities 
are calculated using the NQM and QM as described in the text.
}
\label{fig:exam}
\end{figure}

To illustrate the concept of the linear-regression approach, numerical tests\footnote{The code is implemented in
Java and is included to the package\cite{jhepwork} described in \cite{Chekanov:1261772}.} were 
performed by distributing random points in 2D 
using two Gaussian distributions, see Fig.~\ref{fig:exam}.  
The thick (red) line shows the linear regression which defines the  geometric major axis,
while the thin black line is the geometric minor axis. The eccentricities
were calculated using the NQM and QM.
The following situations were considered: (1) the mean positions
of the Gaussian distributions were set to be the same.
(all eccentricities are close to zero); 
(2) the centers of two Gaussian distributions were shifted by 3 units 
in $\phi$-direction. This leads to non-zero global eccentricities,
and the eccentricities which reflect skewness ($ECC_{MJ}$ and $ECC_{MI}$)
are close to zero. 
When a new  point            
is added with the weight 50 (open circle) (see Fig.~\ref{fig:exam}(3)),
this impacts the values of $ECC_{MI}$.
Moving this point closer to the major axis (Fig.~\ref{fig:exam}(4))
changes the value of $ECC_{MJ}$.

\begin{figure}[htp]
\begin{center}
\mbox{\epsfig{file=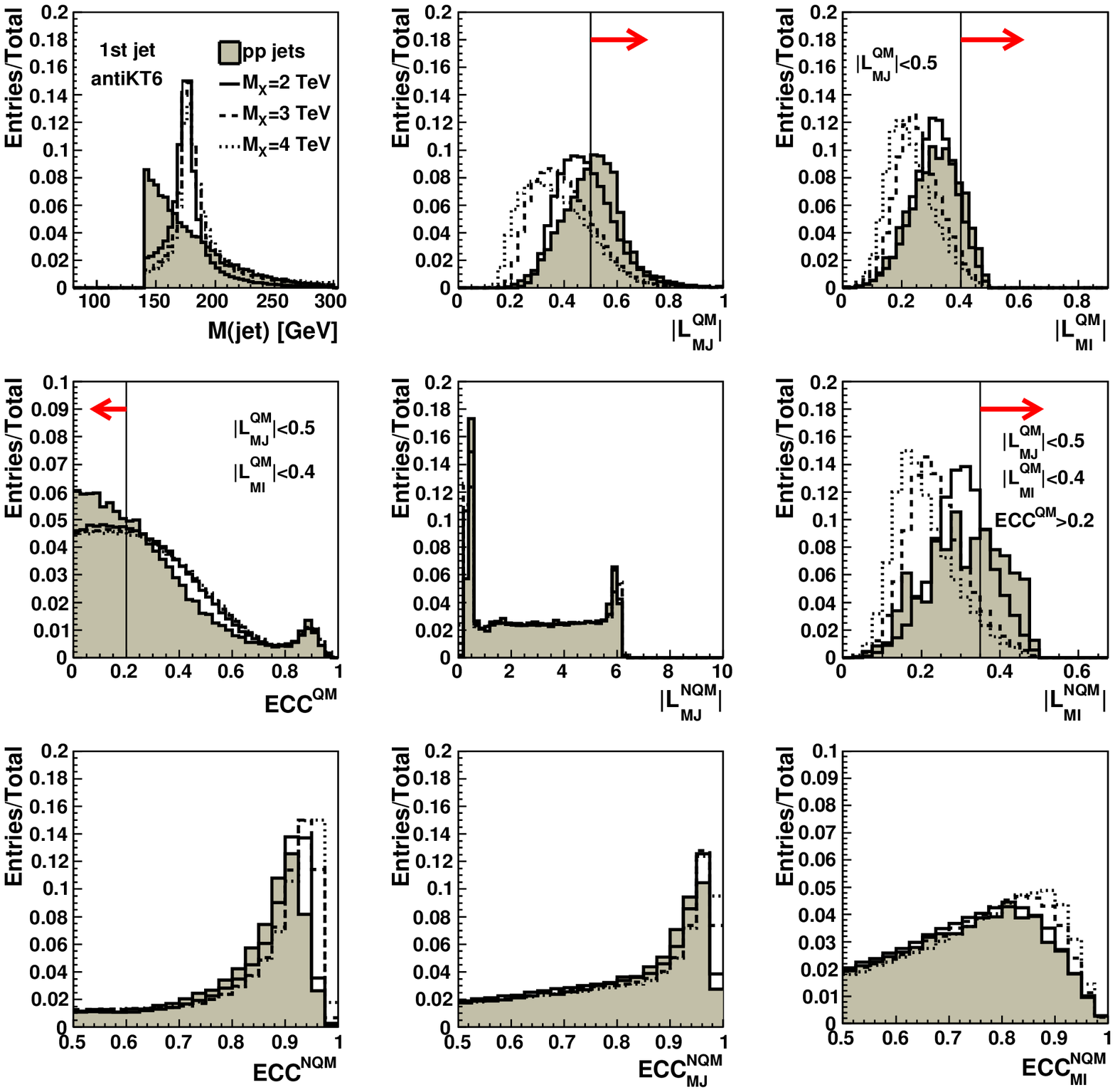,width=15cm}}
\caption
{
Jet mass and jet-shape variables for the leading in $p_T$ jet 
in inclusive
$pp$ collisions (filled histograms) simulated with the PYTHIA model. 
The jet shapes are shown after using the jet-mass cut $M(jet)>140$ GeV.
Also shown are the shape variables
for $X\to t\bar{t} \to W^+b_1 W^-b_2$, with $W$ bosons decaying hadronically into two jets.
The state $X$ was simulated using a $Z'$ particle
with a mass of $2$, $3$ and $4$ TeV (solid and dashed lines, respectively).
Events were selected with at least one jet with $p_T>500$ GeV using the 
anti-$k_T$  jet algorithm.
The vertical lines show the cuts applied to reject inclusive QCD events. 
}
\label{jet1a}
\end{center}
\end{figure}

Figure~\ref{jet1a} shows the variables for the signal and background events for the leading in $p_T$ jet,
after the mass cuts at $140$ GeV as discussed in Sect.~\ref{sec1}. 
For a better shape comparison, all  distributions are normalized
to unity. It should be stressed that, in reality, the cross sections for QCD events can be 
three orders of magnitude larger than those for the signal events\footnote{This statement
is valid for $Z'$ particles included into the PYTHIA predictions.}.
We will discuss this point later; 
at this stage we are only interested in the shape comparison. 
  
\begin{figure}[htp]
\begin{center}
\mbox{\epsfig{file=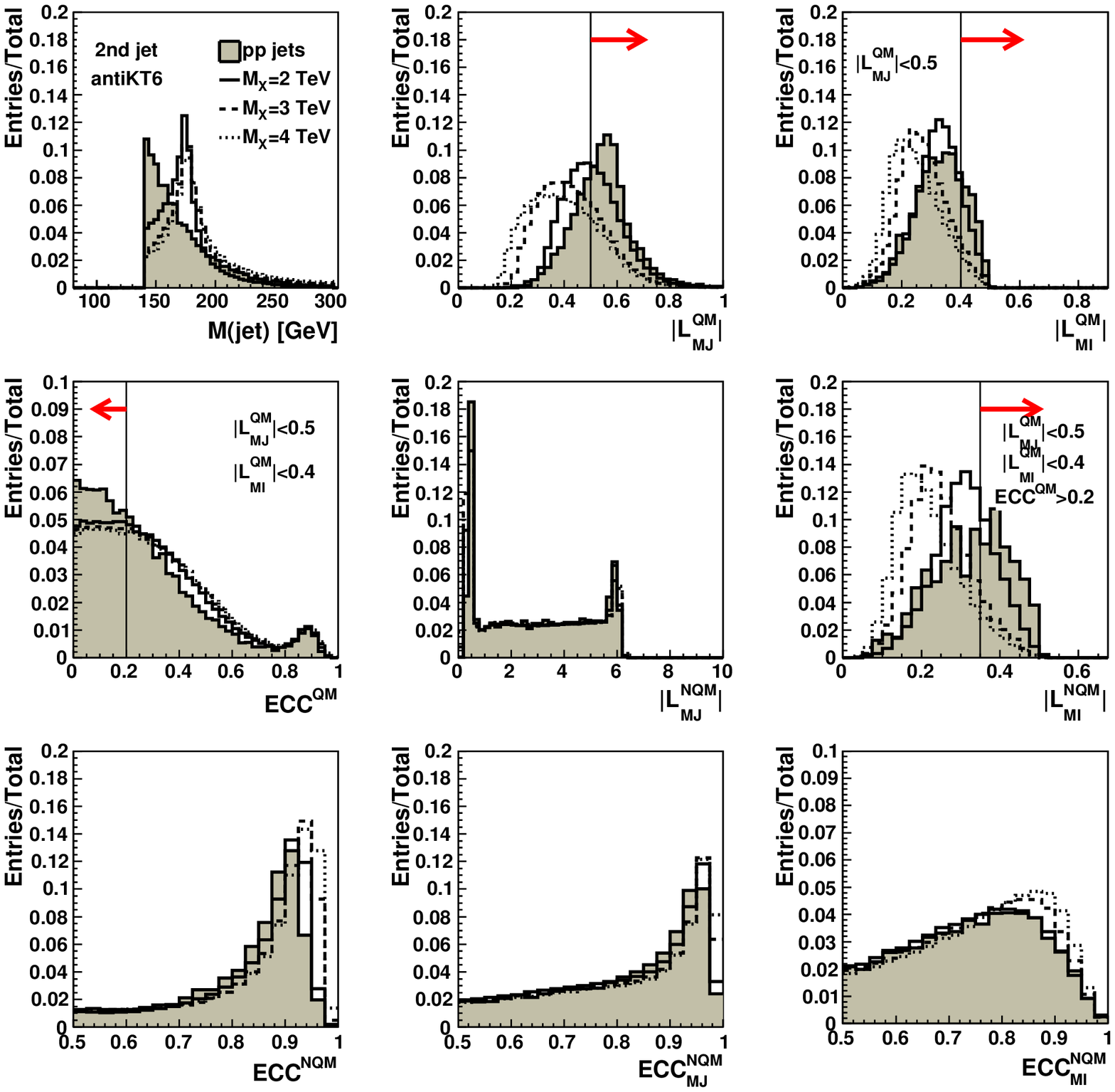,width=15cm}}
\caption
{
Same as Fig.~\protect{\ref{jet1a}}, but for the second leading in $p_T$ jet.
}
\label{jet2a}
\end{center}
\end{figure}

As it can be seen from Figure~\ref{jet1a}, the mass
cut is essential for the TeV-scale particle searches: in addition to the fact that it is strongest
for separation of background events, the differences between jet shapes for the signal and
background events significantly depend on the applied mass cut. 
Similarly, Fig.~\ref{jet2a} shows the same variable but for the second jet. 

Arrows on Figures~\ref{jet1a} and \ref{jet2a} show possible cuts designed to reject QCD events in $pp$ collisions.
The mass cut is applied for all shape distributions. 
It should be noted that  cuts on
the shape variables are applied after using the other cuts (indicated on the figure).
We did not apply the cuts on the eccentricities in the NQM approach since
a cut on the $|\LMI^{NQM}|$ is already sufficient to make sure that only
asymmetric events are accepted.
It can be seen that the jet-shape cuts should be tightened for 3 and 4 TeV particles
to obtain the largest possible rejection for QCD jet events. 

\begin{figure}[htp]
\begin{center}

\centering
\includegraphics[scale=0.48]{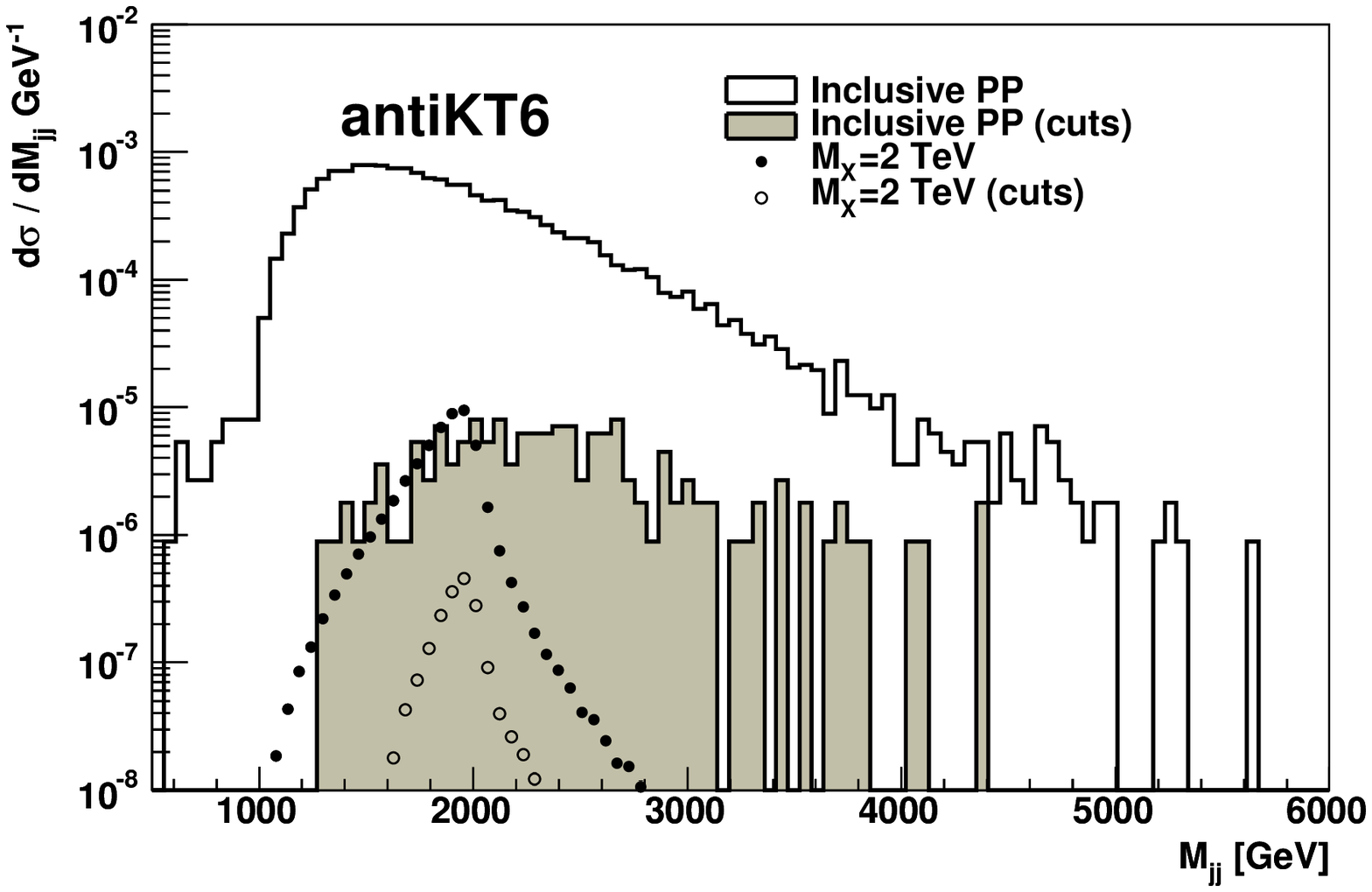}
\centering
\includegraphics[scale=0.48]{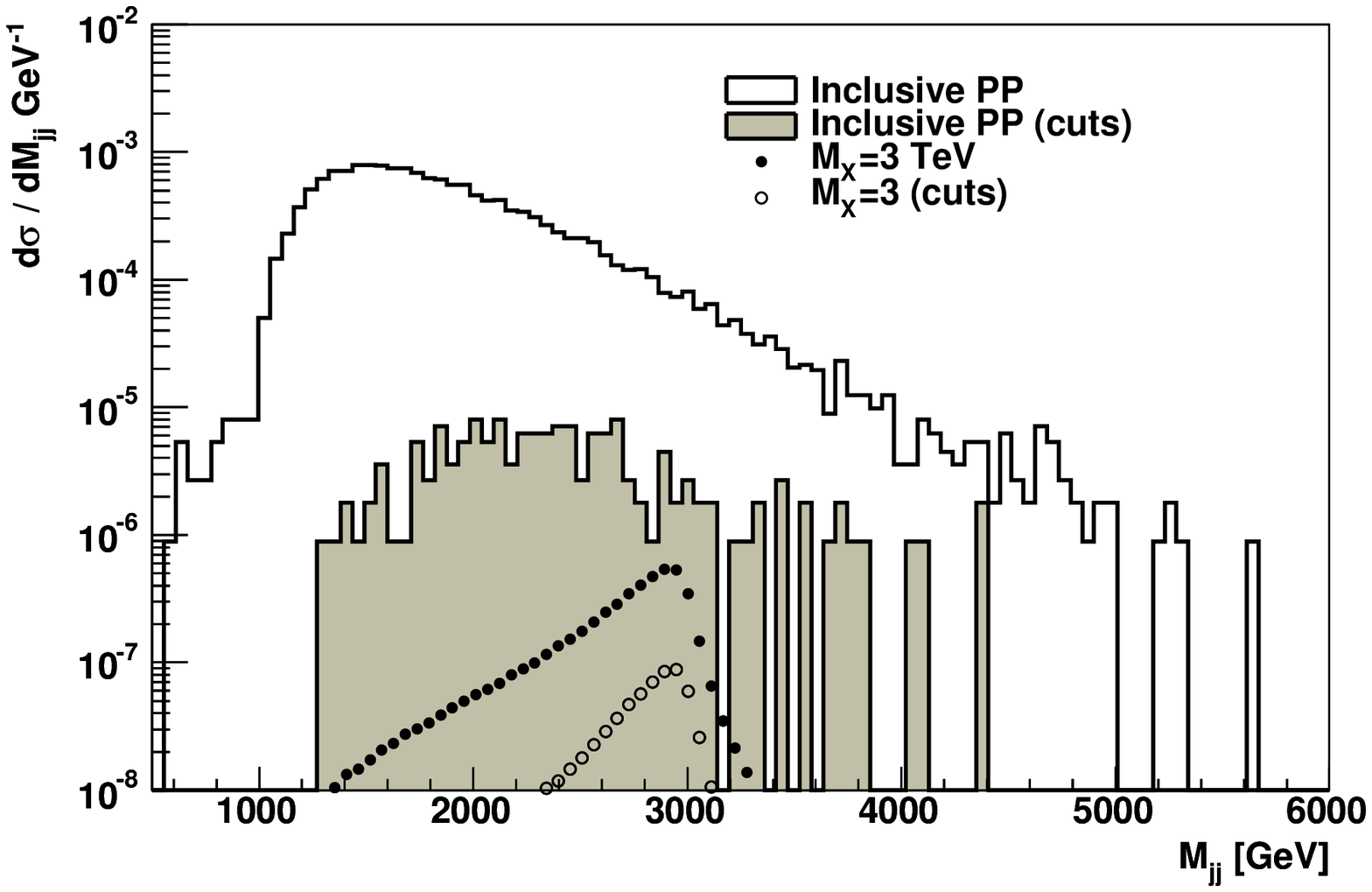}
\centering
\includegraphics[scale=0.48]{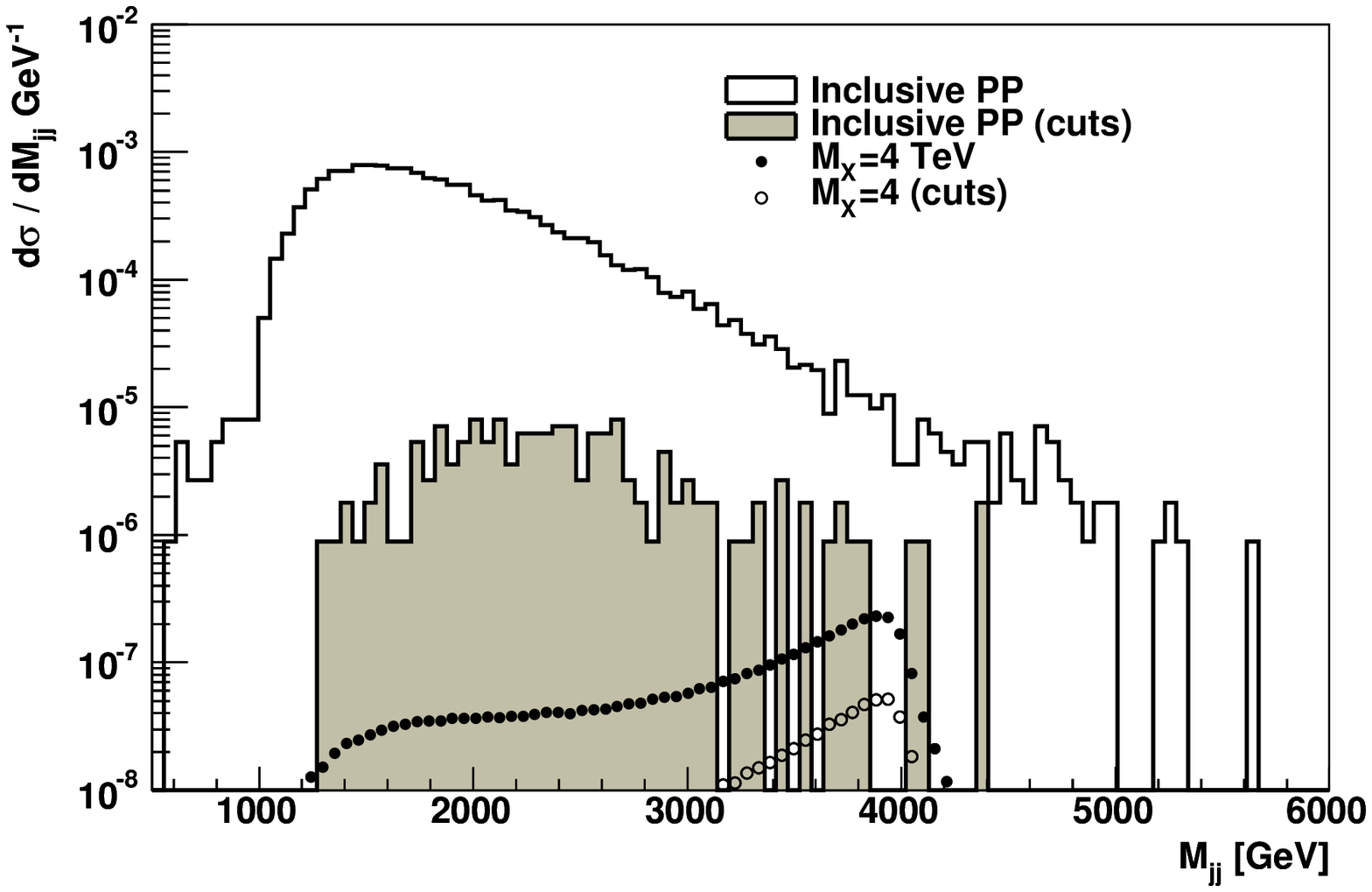}
\caption
{
The differential cross section for the jet-jet invariant mass after the mass cut $M(jet)>140$ GeV 
before (open histograms and filled symbols) and after (filled histogram and open symbols) the jet-shape cuts. 
We used the PYTHIA model for the simulation of $Z'$ particles with 2, 3 and 4 TeV masses. 
The relative size of the signal compared
to the QCD background level increases
after applying the jet-shape cuts.   
See the text for more detailed numerical estimates.
}
\label{jetjet}
\end{center}
\end{figure}

Figure~\ref{jetjet}
shows the expected differential cross sections for the jet-jet invariant mass $M_{jj}$ 
after the mass cut $M(jet)>140$ GeV.
The distributions are shown before and after the applied jet-shape cuts indicated
with the arrays in Fig.~\ref{jet1a} and \ref{jet2a}. 
It can be seen that after the jet-shape cuts, the expected signal (open dots) is a factor of ten
smaller than the QCD background level (the filled histograms), while it is much larger for the jet-mass
cut alone (filled dots and the open histogram). Certainly, the conclusion about 
the relative size of the signal compared to QCD background depends on the underlying model,
which, in this case, was chosen to be the $Z'$. The relative size of the 
signal compared to QCD background
does not change much with increase of  $M_X$, which is mainly due to the fact that no 
readjustments of the jet-shape cuts were done going to  higher masses.

Let us give numerical estimates. The rejection factor $r(QCD)$ for QCD events
in the mass range $1.5-2.5$ TeV
is roughly $100$, while it is a factor $25$ for the 2 TeV signal.
Therefore, the ratio of the rejection factors for inclusive QCD and events with heavy states
is about $3.7$,

\begin{equation}
\frac{r(QCD)}{r(X(2 TeV))} \simeq 3.7. 
\label{eq1}
\end{equation}
The rejection factor for QCD events for $M_{jj}\sim 2$ TeV 
is $44$, while it is only a factor $7.4$ for 3 TeV particles.
Therefore, the ratio of the rejection factors for inclusive QCD and events with 3 TeV states
is larger: 
\begin{equation}
\frac{r(QCD\>\> jets)}{r(X(2\>\> TeV))} \simeq 6.   
\label{eq2}
\end{equation}
For the 4 TeV signal events, the rejection can be as high as 8 after adjusting the cuts
on the shape variables.
Generally, it is expected that the relative rejection factor will be
even larger for higher masses and
roughly follows:

\begin{equation}
\frac{r(QCD\>\>jets)}{r(X(M\>\> TeV))} \propto  M, 
\label{eq3}
\end{equation}
where  $M$ is a mass (in TeV) of  a heavy particle.
At this stage, it is difficult to verify the exact functional dependence on $M$ 
since this depends on the chosen jet-shape  cuts.  
We only can offer an
approximate dependence which qualitatively follows from Figs.~\ref{jet1a} and \ref{jet2a}.

Using the jet-shape rejection rates,
now it is easy to calculate the global rejection factor including the jet-mass
cut. 
We will make our estimates for the 3 TeV exotic particles. According to Table~\ref{table1},
the relative rejection factor for 140 GeV mass cut is 380/1.6=237.
The rejection factor only weakly depends on the mass of  heavy particles after using
the jet-mass cut to consider only jets with masses close to the nominal top mass.
This rejection factor 
should be multiplied by the factor Eq.~(\ref{eq2}) from the use of jet-shape variables.
Thus, the overall relative rejection factor is above 1400. 

For an arbitrary TeV-scale mass $M$ of a heavy state decaying to $t\bar{t}$,
the total relative rejection factor follows this empirical expression:

\begin{equation}
\frac{r_{tot}(QCD\>\>jets)}{r_{tot}(X(M\>\> TeV))} \simeq C \cdot (A+M),  
\label{eq4}
\end{equation}  
where an $C$ is a rejection factor which significantly
depends on the mass cuts as shown in Table~\ref{table1}, 
but relatively independent of the heavy-state mass.
The second factor, $A+M$ (with $A$ being a constant)  originates
from the jet-shape selection and explicitly depends on the mass of a heavy state.

The efficiency of the selection of new states significantly depends on the applied cuts and the
mass of such states. For the example discussed above,
the overal efficiency including the applied mass cuts is roughly $8\%$ 
for a $X$ state with the mass 3 TeV. 

Since the anti-$k_T$ jet algorithm  
turns out to produce circular shapes \cite{Cacciari:2008gp},  it is likely that the use of other jet algorithms
may lead to different rejection factors obtained using the jet shapes.
In particular, we expect that the standard $k_T$ algorithm \cite{Catani:1993hr,*Ellis:1993tq}  
with a larger cone size (0.8-1.0)
will be more suitable for the jet-shape reconstruction.   
It should also be noted that a full detector simulation may change the results.

A comparison of different approaches for QCD background rejection using jet shapes
has been discussed in \cite{Chekanov:2010vc}. 
Usually, a rejection factor 100
for QCD inclusive events is considered as a good starting point for
boosted-object searches in the $t\bar{t}$ channel.
This  rejection heavily depends on the jet-mass cut (the closer the mass cut to the nominal 
top mass, the larger
QCD-event rejection). In this article we have disentangled the mass cut from jet-shape
cuts, showing that a relative jet-shape rejection can be as large as 8 for 4 TeV states, while
the relative rejection factor for QCD events after the mass cut can be above a hundred 
for $M(jet)>140$ GeV (i.e. 380/1.3, see Table.~\ref{table1}).  
Therefore, the overall relative rejection
factor for 4 TeV particles can be close to a thousand.

\section{Conclusions}

The approach proposed in this paper allows to characterize jet shapes beyond the simple
jet-shape characteristics considered in the previous publications 
\cite{Agashe:2006hk,Lillie:2007yh,Butterworth:2007ke,Almeida:2008tp,Almeida:2008yp,
Kaplan:2008ie,Brooijmans:2008,Butterworth:2009qa,Ellis:2009su,ATL-PHYS-PUB-2009-081,CMS-PAS-JME-09-001,Chekanov:2010vc,Almeida:2010pa,Hackstein:2010wk}. In particular,
the current method is 
sensitive to various degrees of skewness of jet shapes in the longitudinal 
(along the major axis) and the transverse (along the minor axis) directions. 
This can be useful for 
searches of $X(\sim TeV)\to t\bar{t}$ states which 
typically have unbalanced jet profiles due to hadronic top decays with the presence of $b$-quark decays.
It was shown that 
the rejection power for QCD jets using the jet-shape characteristics alone can be as high as 8 for 4 TeV
particles for the $X(\sim TeV)\to t\bar{t}$ decay channel. 

It should be noted that this approach is rather
general and can be used for 
any channel with unbalanced energy flows inside a jet due to asymmetric decays.
It can also be used for decays   
where the selection of events with known jet masses
(as in the case of $X\to t\bar{t}$) may not be possible.

\section*{Acknowledgements}
We thank Lily Asquith for discussion and checking alternative jet algorithms. 
The submitted manuscript has been created by UChicago Argonne, LLC, 
Operator of Argonne National Laboratory ("Argonne"). 
Argonne, a U.S. Department of Energy Office of Science laboratory, 
is operated under Contract No. DE-AC02-06CH11357.

\bibliographystyle{./Macros/l4z_pl}
\def\bibname{\Large\bf References}
\def\refname{\Large\bf References}
\pagestyle{plain}
\bibliography{biblio}

\providecommand{\etal}{et al.\xspace}
\providecommand{\coll}{Collaboration}
\catcode`\@=11
\def\@bibitem#1{%
\ifmc@bstsupport
  \mc@iftail{#1}%
    {;\newline\ignorespaces}%
    {\ifmc@first\else.\fi\orig@bibitem{#1}}
  \mc@firstfalse
\else
  \mc@iftail{#1}%
    {\ignorespaces}%
    {\orig@bibitem{#1}}%
\fi}%
\catcode`\@=12
\begin{mcbibliography}{10}

\bibitem{Agashe:2006hk}
K.~Agashe, A.~Belyaev, T.~Krupovnickas, G.~Perez, J.~Virzi,
\newblock Phys. Rev.{} D77~(2008)~015003\relax
\relax
\bibitem{Lillie:2007yh}
B.~Lillie, L.~Randall, L.-T. Wang,
\newblock JHEP{} 09~(2007)~074\relax
\relax
\bibitem{Butterworth:2007ke}
J.~M. Butterworth, J.~R. Ellis, A.~R. Raklev,
\newblock JHEP{} 05~(2007)~033\relax
\relax
\bibitem{Almeida:2008tp}
L.~G. Almeida, S.~J. Lee, G.~Perez, I.~Sung, J.~Virzi,
\newblock Phys. Rev.{} D79~(2009)~074012\relax
\relax
\bibitem{Almeida:2008yp}
L.~G. Almeida, et~al.,
\newblock Phys. Rev.{} D79~(2009)~074017\relax
\relax
\bibitem{Kaplan:2008ie}
D.~E. Kaplan, K.~Rehermann, M.~D. Schwartz, B.~Tweedie,
\newblock Phys. Rev. Lett.{} 101~(2008)~142001\relax
\relax
\bibitem{Brooijmans:2008}
G.~H. Brooijmans,
\newblock {\em {High pT Hadronic Top Quark Identification. Published in "A Les
  Houches Report. Physics at Tev Colliders 2007 -- New Physics Working
  Group"}}, Preprint \mbox{hep-ph/0802.3715}, 2008\relax
\relax
\bibitem{Butterworth:2009qa}
J.~M. Butterworth, et~al.,
\newblock {\em {Discovering baryon-number violating neutralino decays at the
  LHC}}, Preprint \mbox{CERN-PH-TH/2009-073, hep-ph/0906.0728}, 2009\relax
\relax
\bibitem{Ellis:2009su}
S.~D. Ellis, C.~K. Vermilion, J.~R. Walsh,
\newblock Phys. Rev.{} D80~(2009)~051501\relax
\relax
\bibitem{ATL-PHYS-PUB-2009-081}
ATLAS~Collaboration,
\newblock {\em Reconstruction of High Mass $t\overline{t}$ Resonances in the
  Lepton+Jets Channel},
\newblock Technical Report ATL-PHYS-PUB-2009-081. ATL-COM-PHYS-2009-255, CERN,
  Geneva, May 2009\relax
\relax
\bibitem{CMS-PAS-JME-09-001}
CMS~Collaboration,
\newblock {\em A Cambridge-Aachen (C-A) based Jet Algorithm for boosted top-jet
  tagging},
\newblock Technical Report CMS-PAS-JME-09-001, Jul 2009\relax
\relax
\bibitem{Chekanov:2010vc}
S.~Chekanov, J.~Proudfoot,
\newblock Phys. Rev.{} D81~(2010)~114038\relax
\relax
\bibitem{Almeida:2010pa}
L.~G. Almeida, S.~J. Lee, G.~Perez, G.~Sterman, I.~Sung,
\newblock Phys. Rev.{} D82~(2010)~054034\relax
\relax
\bibitem{Hackstein:2010wk}
C.~Hackstein, M.~Spannowsky,
\newblock {\em {Boosting Higgs discovery - the forgotten channel}}, 2010,
  hep-ph:1008.2202\relax
\relax
\bibitem{Sjostrand:2006za}
T.~Sjostrand, S.~Mrenna, P.~Z. Skands,
\newblock JHEP{} 05~(2006)~026\relax
\relax
\bibitem{runmc}
S.~Chekanov,
\newblock Comput. Phys. Commun.{} 173~(2005)~175.
\newblock Available on \texttt{http://projects.hepforge.org/runmc/}\relax
\relax
\bibitem{root}
I.~Antcheva, et~al.,
\newblock Comput. Phys. Commun.{} 180~(2009)~2499\relax
\relax
\bibitem{fastjet}
M.~Cacciari, G.~Salam, G.~Soyez,
\newblock {\em {{FastJet. A C++ library for the $k_T$ algorithm}}},
\newblock available on \texttt{http://www.lpthe.jussieu.fr/\til
  salam/fastjet/}\relax
\relax
\bibitem{Cacciari:2008gp}
M.~Cacciari, G.~P. Salam, G.~Soyez,
\newblock JHEP{} 04~(2008)~063\relax
\relax
\bibitem{Moraes:2009zz}
ATLAS~Collaboration, A.~Moraes,
\newblock {\em {Modeling the underlying event: Generating predictions for the
  LHC}}.
\newblock ATL-PHYS-PROC-2009-045\relax
\relax
\bibitem{jhepwork}
S.~Chekanov,
\newblock {\em {{jHepWork. A multiplatform data-analysis framework}}},
\newblock available on \texttt{http://jwork.org/jhepwork/}\relax
\relax
\bibitem{Chekanov:1261772}
S.~Chekanov,
\newblock {\em Scientific data analysis using Jython Scripting and Java}.
\newblock Springer-Verlag, London, 2010.
\newblock ISBN 978-1-84996-286-5, e-ISBN 978-1-84996-287-2\relax
\relax
\bibitem{Catani:1993hr}
S.~Catani, Y.~L. Dokshitzer, M.~H. Seymour, B.~R. Webber,
\newblock Nucl. Phys.{} B406~(1993)~187\relax
\relax
\bibitem{Ellis:1993tq}
S.~D. Ellis, D.~E. Soper,
\newblock Phys. Rev.{} D48~(1993)~3160\relax
\relax
\end{mcbibliography}

\end{document}